\documentstyle[prl,aps,psfig]{revtex}
\newcommand{\ket}[1]{{|#1\rangle}}
\newcommand{\bra}[1]{{\langle#1|}}
 
\begin{document}

%\preprint{Submitted to Physical Review Letters  		2 April, 1996}

%\draft

\twocolumn[
\vskip -1.5truein
\rightline{LA-UR-96-1266}
\vskip 1.5truein

\title{Decoherence Bounds on Quantum Computation with Trapped Ions}

\author{Richard J. Hughes, Daniel F. V. James, Emanuel H. Knill, 
Raymond Laflamme and Albert G. Petschek}
 
\address{%\vspace*{1.2ex}
 	%\hspace*{0.5ex}{
 	 Los Alamos National Laboratory, Los Alamos, NM 87545, USA \footnote{
Send reprint requests to: Richard J. Hughes, Physics Division P 23, 
Mail Stop H 803, Los Alamos National Laboratory, Los Alamos NM  87545.}%}
        }

\date{\today}
\maketitle

]

%%%%%%%%%%%%%%%%%%%%%%%%%%%%%%%%%%%%%%%%%%%%%%%%%%%%%%%%%%%%%%%%%%%%%%%%%%%%%
% Abstract

\begin{abstract}
Using simple physical arguments we investigate the capabilities of a quantum 
computer based on cold trapped ions.  From the limitations imposed on such a 
device by spontaneous decay, laser phase coherence, ion heating and other sources 
of error, we derive a bound between the number of laser interactions and the number 
of ions that may be used.  The largest number which may be factored using a variety
 of species of ion is determined.  
\end{abstract}

\bigskip 

PACS numbers: 32.80.Pj, 42.50.Vk, 89.80.+h 

\bigskip
% \narrowtext

%%%%%%%%%%%%%%%%%%%%%%%%%%%%%%%%%%%%%%%%%%%%%%%%%%%%%%%%%%%%%%%%%%%%%%%%%%%%%

%Main Text

In a quantum computer binary numbers can be represented by quantum states of 
two-level systems (``qubits''), bringing a new feature to computation: the ability 
to compute with coherent superpositions of numbers \cite{1}.  Because a single quantum
 operation can affect a superposition of many numbers in parallel, a quantum computer
  can efficiently solve certain classes of problems that are currently intractable on classical 
computers, such as the determination of the prime factors of large numbers \cite{2}.  These 
problems are of such importance that there is now considerable interest in the practical
implementation of a quantum computer \cite{3,4}.  There are three principal challenges which
 must be met in the design of such a device: the qubits must be sufficiently isolated from 
the environment so that the coherence of the quantum states can be maintained throughout 
the computation; there must be a method of manipulating the states of the qubits in order 
to effect the logical ``gate'' operations; and there must be a method for reading out the
 answer with high efficiency.  

Cirac and Zoller \cite{5} have made the most promising proposal for the implementation of 
a quantum computer so far.  A number of identical ions are stored and laser cooled in a
 linear radio-frequency quadrupole trap to form a quantum register.  The radio-frequency
 trap potential gives strong confinement of the ions in the $Y$ and $Z$ directions transverse 
to the trap axis, while an electrostatic potential forces the ions to oscillate in an effective 
harmonic potential in the axial direction ($X$).  After laser cooling the ions become
localized  along the trap axis (the Lamb-Dicke regime) with a spacing determined by their
Coulomb  repulsion and the confining axial potential.  The normal mode of the ions'
collective  oscillations which has the lowest frequency is the axial center of mass (CM)
mode, in  which all the trapped ions oscillate together.  A qubit is the electronic ground
state $\ket{g}$($\ket{0}$)  and a long-lived excited state $\ket{e}$($\ket{1}$)  of the
trapped ions.  The electronic configuration of individual  ions, and the quantum state of
their collective CM vibrations can be manipulated by  coherent interactions of the ion with a
laser beam, in a standing wave configuration,  which can be pointed at any of the ions.  The
CM mode of axial vibrations may then be  used as a ``bus'' to implement the quantum logical
gates.  Once the quantum computation  has been completed, the readout is performed through
the mechanism of quantum jumps.   Several features of this scheme have been demonstrated
experimentally, mostly using a {\em single} trapped ion \cite{4,6}.

The unavoidable interaction of a quantum computer with its environment places 
considerable limitations on the capabilities of such devices \cite{7}.  In this letter we 
make a quantitative assessment of these limitations for a computer based on the 
Cirac-Zoller cold-trapped-ion design, in order to determine the best physical 
implementation and the optimization parameters for quantum algorithms \cite{8}.  
There are two fundamentally different types of decoherence during a computation: 
the intrinsic limitation imposed by spontaneous decay of the metastable states $\ket{e}$ of 
the ions; and practical limitations such as the random phase fluctuations of the laser 
driving the computational transitions or the heating of the ions' vibrational motion.  
One could, in principle, expect that as experimental techniques are refined, the effects
 of these practical limitations may be reduced until the intrinsic limit of computational 
capability due to spontaneous emission is attained.

	The influence of spontaneous emission on a quantum computation with trapped 
ions depends on: the natural lifetime of the $\ket{e}$ qubit; the number of ions, $L$, being
used;  and the quantum states of those ions.  The number of ions which are not in their ground
 states varies as the calculation progresses, with ancillary ions being introduced and 
removed from the computation.  The progression of the ions' states can be characterized
 well by an effective number of ions, $L_e$, which have a non-zero population in the excited 
state $\ket{e}$.  In the case of Shor's factoring algorithm \cite{2} (using long
multiplication), a  reasonable estimate is $L_e\approx 2L/3$.  Therefore, to estimate the
effect of decoherence during the  implementation of Shor's algorithm, we will consider the
following process: a series  of laser pulses of appropriate strength and duration (
$\pi/2$ pulses) is applied to $2L/3$ ions, causing  each of them to be excited into an equal
superposition state $(\ket{g}+\ket{e})/2$.  After an interval $T$, a second  series of laser
pulses ($-\pi/2$  pulses) is applied, which, had there been no spontaneous emission, would
cause each ion to be returned to its ground state.  This is the ``correct'' result of our 
pseudo-computation.  If there were spontaneous emission from one or more of the ions,
 then the ions would finish in some other, ``incorrect'' state.  This process involves the
sort  of superposition states that will occur during a typical quantum computation, and so the
 analysis of decoherence effects in this procedure will give some insight into how such 
effects influence a real computation.  A simple calculation shows that the probability of 
obtaining a correct result is 
\begin{equation}
P(T) \approx 1-LT/6\tau_o
\label{pofdecay}
\end{equation}
where  is the natural lifetime of the excited state $\ket{e}$.  Thus the effective coherence
time of  the computer is $6\tau_o/L$.

	The total time taken to complete a calculation will be approximately 
equal to the number of laser pulses required multiplied by the duration of each pulse.  
The time taken to switch the laser beam from ion to ion is assumed to be negligible.  
There are two types of laser pulse that are required in order to realize Cirac and Zoller's 
scheme.  The first requires pulses that are tuned precisely to the resonance frequency of 
the  $\ket{e}$ to $\ket{g}$ transition, configured so that the ion lies at the node of the
laser standing wave  (``V-pulses''); the second requires pulses tuned to the CM phonon
sideband of the  transition, arranged so that the ion lies at the antinode of the standing
wave  (``U-pulses'') \cite{9}.  The interaction of U-pulses with the ions is considerably
weaker  than the V-pulses, and so, assuming constant laser intensity, the U-pulse duration 
must be longer.  Hence, in calculating the total time required to perform a quantum 
computation, we will neglect the time required for the V-pulses.  Because the entire 
calculation must be performed in a time less than the coherence time of the computer, 
we obtain the following inequality:
\begin{equation}
N_Ut_U < 6\tau_o/L
\,,
\label{nutu}
\end{equation}
where  is the total number of U-pulses, each of which has duration $t_u$.  The Hamiltonian 
for the interaction of these pulses with the ions is given by the following expression 
(ref.\cite{5}, eq. (\ref{pofdecay})):
\begin{equation}
{\hat H} = \frac{\hbar\eta}{2\sqrt{L}} \Omega 
 [\ket{e}\bra{g}\hat a e^{-i\phi}  + \ket{g}\bra{e}\hat a^\dagger e^{i\phi} ]
\label{hamiltonian}
\end{equation}
In this formula,  is the Rabi frequency for the laser-ion interaction, L is the number of 
ions in the trap,  $\hat a$ ($\hat a^\dagger$) is the annihilation (creation) operator for
phonons of the CM mode  and  $\eta=\sqrt{\hbar\omega^2\cos^2\theta /2Mc^2\nu_x}$ is the
Lamb-Dicke parameter (here $\omega$ is the laser angular frequency, $\theta$ the angle 
between the laser and the trap axis, $\nu_x$ is the angular frequency of the ions' axial CM
mode  and $M$ the mass of each ion).  A careful calculation, based on a perturbative analysis
of  the excitation of phonon modes other than the CM mode, shows that this Hamiltonian is 
valid if  $(\Omega\eta/2\nu_x\sqrt{L})^2\ll 1$.  The longest duration laser pulse that will
be required to implement a quantum  computing algorithm using a Cirac-Zoller quantum computer
is a U-pulse of duration $t_U=2\pi\sqrt{L}/\Omega\eta$.   We will assume that all  of the
U-pulses required for the calculation are of this duration.   Therefore the lower bound on the
duration of laser pulses is $t_U=y\pi/\nu_x$, where $y$ is a dimensionless  ``safety
factor''.  This result can also be obtained from the naive uncertainty principle  argument
that there must not be appreciable power at the frequencies of the adjacent  lattice
vibrations.

	In order to attain the highest possible computational capability, one will need 
to minimize the duration of each laser pulse.  Hence, it will be advantageous to employ 
an ion trap with the largest possible value of the trap frequency $\nu_x$.  However, the
axial  frequency cannot be made arbitrarily large because, in order to avoid crosstalk
between  adjacent ions, the minimum inter-ion spacing must be much larger than the size of
the  focal spot of the laser beam.  The minimum separation distance between two ions occurs 
at the center of the string of ions, which can be calculated by solving for the equilibrium 
positions of the ions numerically, resulting in the following expression :
\begin{equation}
x_{\rm min}\cong \left({Z^2e^2\over4\pi\epsilon_o\nu_x^2M}\right)^{/13} {2.0\over L^{0.56}}
\,,
\label{xmin}
\end{equation}
where $Z$ is the degree of ionization of the ions, $e$ is the electron charge and  
$\epsilon_o$ is the 
permittivity of a vacuum.  The spatial distribution of light in focal regions is well 
known \cite{10}.  The approximate diameter of the focal spot is $x_{\rm spot} \approx
\lambda F$, where $\lambda$ is the laser  wavelength and $F$ the f-number of the focusing
system (i.e. the ratio of the focal length
 to the diameter of the exit pupil).  Hence the requirement that the ion separation must be 
large enough to avoid cross-talk between ions, i.e. that $x_{\rm min} \gg x_{\rm spot}$,
leads to the following  expression for the duration of the U-pulses:
\begin{equation}
t_U\equiv {\pi y\over \nu_x} = 2.9[s\, m^{-3/2}] \sqrt{Ay^5\lambda^3F^3L^{1.68}\over Z^2}
\,.
\label{tu}
\end{equation}
where $A$ is the atomic mass number of the ions.  From eqs. \ref{nutu} and \ref{tu} we obtain
the following constraint on the number of ions L and the total number of U-pulses :
\begin{equation}
N_uL^{1.84}<2.0[s^{-1}\, m^{3/2}] \frac{Z\tau_o}{y^{5/2} A^{1/2} F^{3/2} \lambda^{3/2}}
\,.
\label{nul}
\end{equation}
	We will now apply this bound to Shor's factor finding algorithm \cite{2}.  Let  be 
the number of bits of the integer we wish to factor.  A careful analysis of the implementation
 of the algorithm (using long multiplication) reveals that the required number of ions and 
U-pulses are given by:
\begin{equation}
L=5l+2 
\,, 
\label{l}
\end{equation}
\begin{equation}
N_U=544l^3+78l^2+10l
\,.
\label{nufunctionl}
\end{equation}
Note that there are asymptotically much more efficient implementations, but they do 
not become competitive for the small number of binary digits under consideration here.  
If the measured Fourier transform \cite{11} and interleaving measurements were to be used 
in the computation, the number of ions required can be reduced to $3l+4$.  However the 
intermediate measurements may increase the decoherence of the other ions due to 
scattered photons or unintended heating of the ions.  It is for this reason that we have 
avoided use of this technique in the assumptions underlying the algorithm. 

	Equations \ref{l} and \ref{nufunctionl} define a curve in $(L,N_U)$ space, which taken in
conjunction  with the inequality (\ref{nul}) allow us to determine the largest number of ions
that can be used  to implement Shor's algorithm in an ion trap computer with bounded loss of
coherence.   The linear relationship between $L$ and $l$, eq.(\ref{l}), can then be used to
determine the largest  number that can be factored.  

	As  specific examples, we will consider the intrinsic computational capacity of 
Cirac-Zoller quantum computers based on the following three ions:

\noindent (i) Hg II: Z=1, A = 198; $\ket{e}$ is a sublevel of the 
$5{\rm d}^9 6{\rm s}^2 {\ }^2{\rm D}_{5/2}$ state,
$\ket{g}$  is the $5{\rm d}^{10} 6{\rm s}^2 {\ }^2{\rm S}_{1/2}$,
the two states being connected  by an electric quadrupole transition: $\lambda$ = 281.5 nm;
$\tau_o\approx 0.1s$.  
 
\noindent (ii) Ca II: Z=1, A = 40; $\ket{e}$ is a sublevel of the 
$3{\rm d}\,{\ }^2{\rm D}_{5/2}$ state,  
$\ket{g}$ is the $4{\rm s}\, {\ }^2{\rm S}_{1/2}$, the
two states being connected  by an electric quadrupole transition: $\lambda$ = 729 nm;
$\tau\approx 1.14s$. 

\noindent (iii) Ba II: Z=1, A = 137; $\ket{e}$ is a sublevel of the
$5{\rm d}\, {\ }^2{\rm D}_{5/2}$ state,  $\ket{g}$ is the $6{\rm s}\, {\ }^2{\rm S}_{1/2}$,
the two states being connected  by an electric quadrupole transition: $\lambda$ = 1.76 $\mu$m; 
$\tau_o\approx 47 s$.

\noindent We shall assume that we have a very high numerical aperture focusing system, so
that   (although in practice such a high focal ratio would be difficult to achieve), and we
will  err on the side of optimism by putting the safety factor $y= 1$.  In figure 1 we have
plotted  the curves which limit the allowed values of L and , as given by eq.(\ref{nul}). We
have also  plotted, with a solid line, the ``curve of factorization'' defined by eqs.
(\ref{l}) and (\ref{nufunctionl}).  The  interception of the limiting curves for the
different ions with the curve of factorization gives
 us the largest allowed value for the number of ions.  Examining these curves, we find 
that the size of the largest integer that can be factored by a Cirac-Zoller quantum computer 
based on Hg II, Ca II or Ba II ions is 6 bits, 9 bits and 13 bits respectively.  If these 
calculations were repeated with the less optimistic value for the safety factor, $y = 3$, one 
obtains 3 bits, 5 bits and 7 bits for the three species of ions, respectively.  We note that 
the spontaneous emission lifetime is proportional to an odd power of $\lambda$ ( 
$\lambda^5$ in the case of 
electric quadrupole transitions) and so, for greater capability, eq.(\ref{nul}) suggests
either  going to longer wavelength (as seen with the three ion species above) or more highly 
forbidden transitions.

\begin{figure}[tbp]
\begin{center}
	\mbox{\psfig{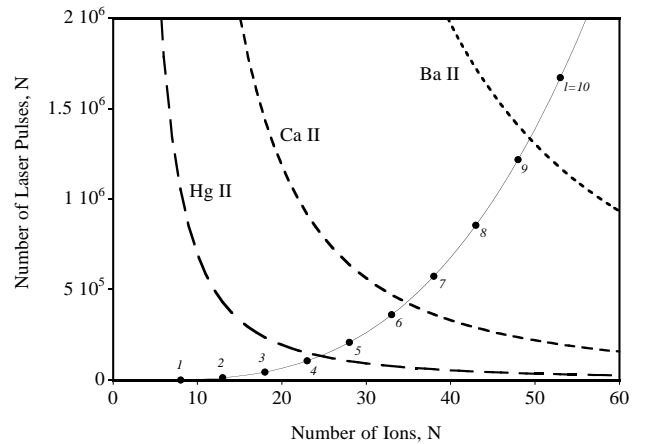}}
\end{center}
\caption{The bounds on the numbers of ions, L, and the number of U-pulses, $N_U$ , that may 
be used in a quantum computation without loss of coherence.  The allowed values of $N_U$ and L 
lie to the left of the curves.  Curves for three ions are plotted.   The unbroken line is the 
``factorization curve'', specified by eqs.(7) and (8), which represents those values of L
and   which are required for execution of Shor's algorithm; the heavy black dots on this line 
represent the values of L and  required to factor a number of l bits (l = 1, 2, ...15).}
\label{fig:bounds}
\end{figure}

	As a more dramatic illustration of the theoretical possibilities of the Cirac-Zoller 
scheme, one may consider a computer based on the  
$4{\rm f}^{14}6{\rm s}\, {\ }^2{\rm S}_{1/2} \leftrightarrow 
4{\rm f}^{13}6{\rm s}^2\, {\ }^2{\rm F}_{1/2}$
electric octupole transition of
Yb II.   This very long lived transition, which has received considerable attention because of
its  potential applications as an optical frequency standard, has a wavelength of 467 nm and 
a calculated lifetime of 1533 days \cite{12}.  Performing a similar calculation to that given 
above suggests that, using this ion, it might be possible to factor a 438-bit number.  
Because such a calculation would require around 2200 trapped ions and  $4.5\times 10^{10}$ U-pulses, 
taking about 100 hours, it would be difficult to over-emphasize the problems attendant 
on such an experiment.

	One may calculate the limits on factoring due to other causes of decoherence by 
a similar procedure to that used above.  In this case, we will assume that the loss of 
quantum coherence due to sundry effects such as random fluctuations of the laser phase
 or the heating of the ions' vibrational motion can be characterized by a single coherence 
time $\tau_e$.  The effects of other causes of error, such as imprecise measurement of the
areas  of $\pi$-pulses, which do not result in decoherence but nevertheless lead to incorrect
results in a computation, can also be characterized by the time $\tau_e$.  Thus, in place of
eq.(\ref{nutu}) we  now have the inequality .  Using eq.(\ref{tu}) we obtain the following
constraint on the values of the number of ions L and the number of laser pulses  which can be
used in a factoring experiment without significant loss of quantum coherence:
\begin{equation}
N_uL^{0.84}<0.34[s^{-1}\, m^{3/2}] \frac{Z\tau_e}{y^{5/2} A^{1/2} F^{3/2} \lambda^{3/2}}
\,, 
\label{nul2}
\end{equation}
Using the ``factorization curve'' specified by eqs. (7) and (8), one can obtain as
before a  value for the number of bits $l$ in the largest number which may be factored.  In
this case  the value of $l$ will depend on the value of the coherence time $\tau_e$.  In
figure 2 we have plotted  the values of $l$ as a function of the experimental coherence time
for the three species of  ions discussed above.  As $\tau_e$ increases, the largest number
that can be factored also increases,  until the limit due to spontaneous emission discussed
above is attained.  The slowest heating  rate for a single trapped ion so far reported is 6
phonons per second (i.e. $\tau_e=0.17 s$)\cite{13}, and the  laser phase coherence times
longer than $10^{-3}s$ have been achieved by several groups
\cite{14}.   Comparing these numbers with fig.2, we see that, in principle, current technology
is  capable of producing a quantum computer that could factor at least small numbers 
(several bits).  Note that, in contrast to the spontaneous emission bounds from eq. (6) 
(where $\tau_o$  is, for quadrupole transitions, proportional to $\lambda^5$), eq.(9) argues
for using shorter wavelength transitions.  So we see from figs. 1 and 2 that Ca II is a good
choice of ion for the experimental study of this technology because it allows a large number
of  operations to be performed with realistic laser stability and ion heating requirements.

	The various causes of experimental decoherence which are mentioned above 
are all the subject of on-going research.  It is not clear, for example, how laser phase 
fluctuations will affect quantum computations; it may be the case that the laser need 
be coherent only over the period required to execute each quantum gate operation.  
Furthermore, the heating rate of the ions' vibrational motion as a function of the 
number of trapped ions is not known.  Other methods of coherent population transfer, 
which may be less susceptible to the effects of phase fluctuations, for example 
stimulated Raman adiabatic passage (STIRAP) \cite{15}, are being investigated.  

\begin{figure}[tbp]
\begin{center}
	\mbox{\psfig{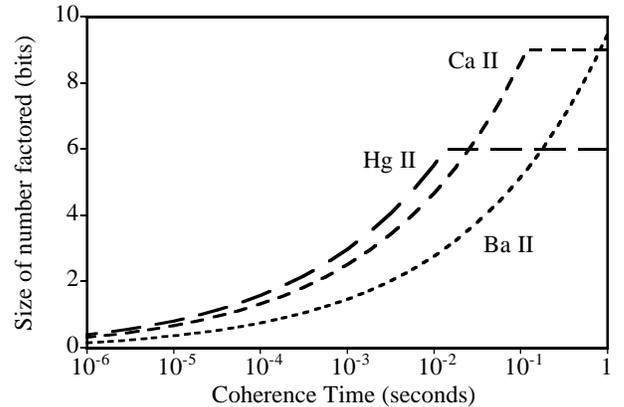}}
\end{center}
\caption{The variation of the number of bits l  in the largest integer that may be factored 
with the experimental coherence time for the three ions discussed in the text.  The maximum 
values of the computational capacities for the ions Hg II and Ca II are the limits determined 
by spontaneous emission.}
\label{fig:cohtime}
\end{figure}

	It is clear that if quantum computation is to overcome decoherence and other 
errors, then error correction must be used extensively.  So far, suggestions for error 
correction have relied either on variations of the ``watch dog'' effect \cite{16,17}, or on 
exploiting the properties of certain entangled states to reduce the impact of decoherence 
in a quantum memory \cite{18}.  The latter has not yet proven to be practical for use during 
a quantum computation, primarily because there has not been any analysis of the success 
of the method under realistic assumptions on operator errors.  If operational errors were
 negligible, the effect of decoherence on quantum memories could be reduced arbitrarily.  
However some of the ``watch dog'' methods that have been suggested are quite practical.  
For example, many computations require the use of ancillary qubits which are periodically 
returned to the ground state.  Measuring these ancillas when they are supposed to be in 
the ground state can be used to help dissipate errors.  Recent simulations \cite{17} indicate 
that this method is indeed helpful in maintaining the state of the computation.  
Implementation of the method does require intermediate measurements.   In any 
case the effect of using a ``watch dog'' method is to stabilize the effective decoherence 
time $\tau_e$ by ensuring less dependence between the errors of successive operations.

	In conclusion, we have derived quantitative bounds which show 
how the computational capabilities of a trapped ion quantum computer depend 
on the relevant physical parameters and determine the computational ``space'' ($L$) 
and ``time'' ($N_U$) combination that should be optimized for the most effective algorithms.  
The effect of this bound has been illustrated by calculating the size of the largest number 
that may be factored using a computer based on various species of ion.  Our results 
show there is reason for cautious optimism about the possibility of factoring at least 
small numbers using a first generation quantum computer design based on cold trapped 
ions.  However, the large number of precise laser operations required and the number of 
ions involved indicates that even this computationally modest goal will be extremely 
challenging experimentally.  

The authors would like to thank Rainer Blatt, Ignacio Cirac, Heidi Fearn, Hugh Klein, 
Norman Kurnit, Stephen Lea and Chris Monroe for helpful advice and useful comments.  
This research was funded by the National Security Agency.


\begin{thebibliography}{99}
\bibitem{9} As we are considering quadrupole transitions, rather than the dipole transitions 
discussed in ref. \cite{5}, the roles of the nodes and antinodes of the standing wave are
reversed.   We have used the notation ``U'' and ``V'' to denote the pulses which implement the
required  unitary transformations as defined in \cite{5}.
\end{thebibliography}
\end{document}